\documentclass[american,aps,pra,reprint,superscriptaddress,longbibliography]{revtex4-1}
\usepackage{amsmath,amssymb,graphicx, bm}
\usepackage[unicode=true,pdfusetitle, bookmarks=true,bookmarksnumbered=false,bookmarksopen=false, breaklinks=false,pdfborder={0 0 0},backref=false,colorlinks=false] {hyperref}
\hypersetup{colorlinks,linkcolor=myurlcolor,citecolor=myurlcolor,urlcolor=myurlcolor}
\usepackage{braket,colortbl,amsthm,amsmath,amssymb,txfonts,graphicx}
\definecolor{myurlcolor}{rgb}{0,0,0.7}
\usepackage{float}
\usepackage{cleveref}
\usepackage{xcolor}
\usepackage{subfigure}

\theoremstyle{plain}
\usepackage{mathrsfs}
\usepackage{comment}
\usepackage{ragged2e}  % for \justifying

\usepackage[utf8]{inputenc}
\usepackage[normalem]{ulem}
\usepackage{float}
\usepackage{mathrsfs}
\usepackage{braket}
\usepackage{physics}
\usepackage{mwe}
\usepackage{pifont}% http://ctan.org/pkg/pifont

\begin{document}

\title{Quantum Mpemba Speedups in the Thermodynamics of Landauer Erasure} 

\author{Pritam Chattopadhyay}
\email{pritam.chattopadhyay@weizmann.ac.il}
\affiliation{Department of Chemical and Biological Physics, Weizmann Institute of Science, Rehovot 7610001, Israel}

\date{\today}

\begin{abstract}
We investigate how nonequilibrium quantum initial states can reduce the finite-time thermodynamic cost of Landauer erasure. Considering a general finite-dimensional quantum memory coupled to a thermal reservoir via a Davies generator, we show that the dissipated heat at a fixed operational erasure fidelity is largely controlled by the overlap of the initial state with the slowest Liouvillian relaxation mode. We derive a modified finite-time Landauer bound in which the excess dissipation above the quasistatic limit scales quadratically with this slow-mode projection, and we prove a sufficient \textit{Mpemba–Landauer} condition under which a hotter state can erase faster and dissipate less heat than a colder preparation, without \textit{violating} Landauer’s principle. A minimal qutrit model illustrates these \textit{quantum Mpemba speedups} and reveals broad parameter regimes where coherence and Hamiltonian-induced modes conspire to suppress finite-time entropy production. We further identify practical control knobs, including temperature tuning, coherence engineering, and Hamiltonian shaping of Liouvillian spectra, which enable Mpemba-enhanced erasure to be directly tested on platforms such as superconducting circuits, trapped ions, semiconductor quantum dots, and solid-state spins. 
\end{abstract}

\maketitle

\section{Introduction}
Landauer’s principle (LP), a.k.a, Landauer erasure (LE)~\cite{landauer1961irreversibility,landauer1991information,goold2015nonequilibrium,lorenzo2015landauer,proesmans2020finite,mandal2012work,deffner2013information,PhysRevLett.107.010604,moore2012landauer,chattopadhyay2025landauer,bennett1982thermodynamics,buffoni2022spontaneous,reeb2014improved,LutzRVW,lubkin1987,chattopadhyay2025understanding1,tang2025information} establishes a fundamental connection between information processing and thermodynamics: any logically irreversible erasure of a single bit requires at least $k_{B}T \ln 2$ of dissipated heat in the quasistatic limit. Realistic erasure protocols, however, operate in finite time and therefore incur additional dissipation arising from nonequilibrium relaxation~\cite{goold2015nonequilibrium,lorenzo2015landauer,proesmans2020finite}. Understanding how to suppress this excess thermodynamic cost has become a central challenge for nanoscale devices~\cite{yan2018single,an2015experimental,huber2008employing,rossnagel2014nanoscale,saira2020nonequilibrium}, quantum information platforms~\cite{mandal2012work,buffoni2023cooperative,PhysRevLett.134.100401,PhysRevE.109.024138,deffner2013information}, and the emerging thermodynamics of computation~\cite{ray2023gigahertz,chu2018thermodynamically,strasberg2015thermodynamics,PRXEnergy.4.023008}.

In parallel, recent advances in nonequilibrium statistical mechanics have uncovered surprising anomalies in relaxation dynamics, most prominently the \textit{ Mpemba effect}~\cite{Mpemba1969,Lasanta2017,Campisi2021Mpemba,Lu2017,Klich2019,Bechhoefer2021,Zhang2020,Kumar2020,Pal2021,moroder2024thermodynamics,teza2025speedups,carollo2021exponentially,ares2025quantum,Yu2025QuantumMpembaSymmetry,Warring2024ExploringQME,PhysRevLett.134.107101,PhysRevLett.133.136302,PhysRevLett.133.010402,PhysRevLett.133.140405,PhysRevLett.133.010401,wgr5lb6b,zhang2025observation,PhysRevX.9.021060,longhi2025mpemba,alyuruk2025thermodynamic,PhysRevLett.131.080402,g94p-7421,52y58kl2,chattopadhyay2026anomaly,PhysRevLett.134.220403,biswas2023mpemba,li2025canonical,5xrrx2rm,chatterjee2025direct,schnepper2025experimental,mondal2025mpemba,bagui2025detection,hallam2025tunable,zhang2026engineering,song2026quantum,peluso2026optimal,rocha2026description,zeng2026theory,liu2026stronger,benjadi2026exponential,Bao2026Entanglement} where a \textit{hotter} initial state can relax to \emph{equilibrium faster} than a \textit{colder} one. Quantum generalizations of this phenomenon~\cite{moroder2024thermodynamics,PhysRevLett.133.136302} have shown that such inversions originate from the spectral structure of the Liouvillian generator governing dissipative evolution. Specifically, the projection of the initial state onto the dominant slow relaxation mode determines the effective relaxation rate, creating the possibility that \textit{high-temperature} or \textit{coherently prepared states} may bypass the slowest dynamical channel.

\begin{figure}[htpb]
    \centering
    \includegraphics[width=0.95\linewidth]{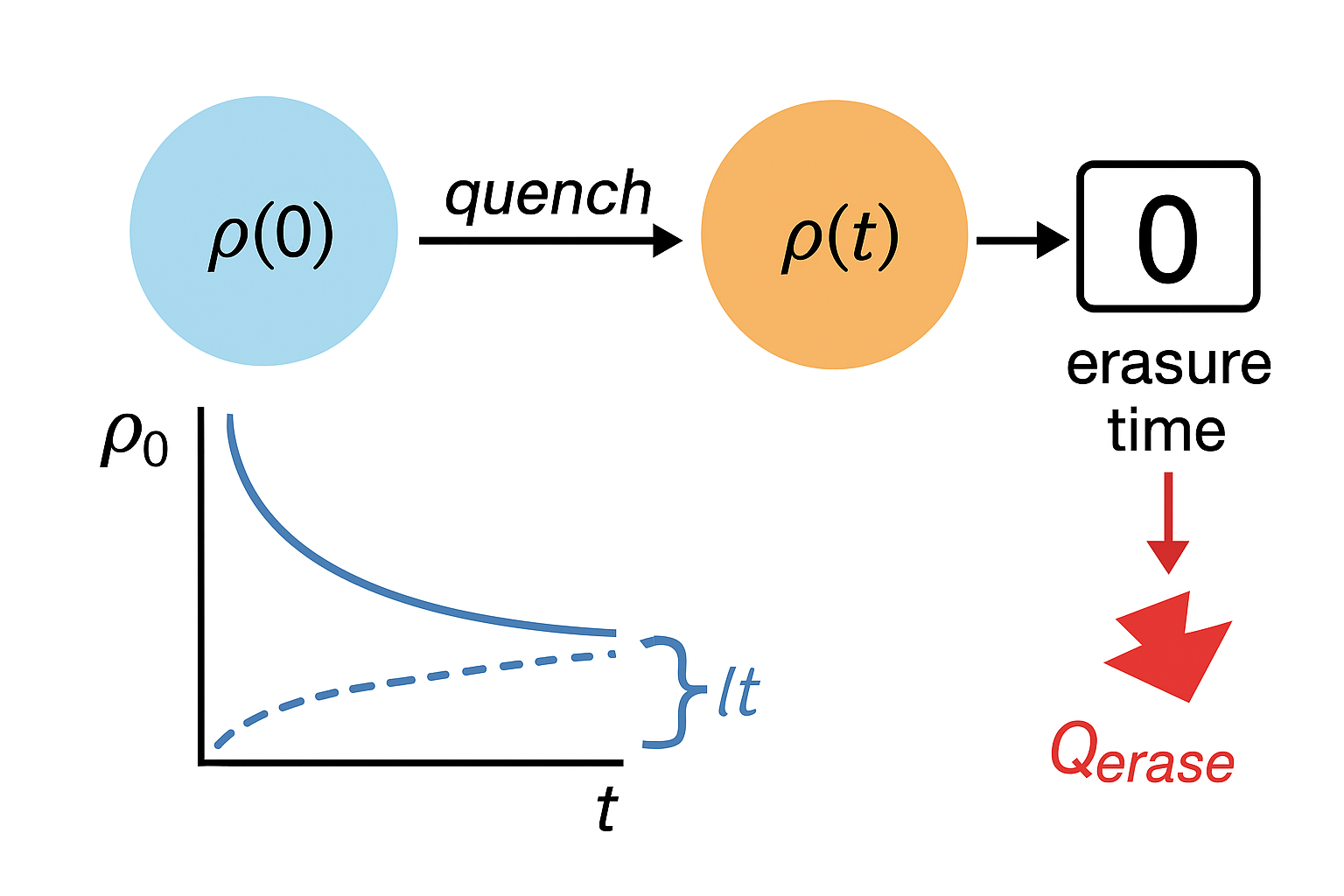}
    \caption{A general setup of quantum Mpemba–enhanced Landauer erasure. A quantum memory is encoded in a finite-dimensional Hilbert space %$\mathcal{H}$, 
    with logical-zero subspace, %$\mathcal{H}_0$ (projector $\Pi_0$) 
    and logical-one is stored in the orthogonal complement. The system is initially prepared in a thermal state $\rho_{\text{th}}(T_{\text{prep}})$ of an input Hamiltonian, then quenched to an erasure Hamiltonian and weakly coupled to a bath at temperature $T_b$, generating Lindblad dynamics that drive the memory toward the logical zero subspace. 
    }
    \label{fig:placeholder}
\end{figure}

In this work, we build upon the established Liouvillian spectral understanding of the \textit{quantum Mpemba effect} (QME) and investigate its implications for the \textit{finite-time thermodynamics of LE}. Rather than proposing a new microscopic mechanism for the Mpemba effect, we show that its well-known spectral origin can be harnessed as a thermodynamic resource for information erasure. Working in a general finite-dimensional quantum setting, we demonstrate that initial thermal states with reduced overlap with the slowest erasure-relevant Liouvillian mode reach a prescribed logical-erasure fidelity in a shorter operational time. We further derive conditions under which this dynamical speedup translates into reduced finite-time entropy production and a modified \textit{finite-time Landauer bound (LB)}, while remaining fully consistent with the laws of thermodynamics. Our results, therefore, connect the established spectral theory of the QME with the operational setting of finite-time information erasure, identifying nonequilibrium state preparation for improving the energetic efficiency of quantum memories.

\section{General setup}
We consider a quantum memory (Fig.~\ref{fig:placeholder}) encoded in a finite-dimensional Hilbert space $\mathcal H$ of dimension $d$. 
The logical state $0$ is stored in a designated subspace $\mathcal H_{0}$, described by a projector $\Pi_{0}$, while logical $1$ occupies the orthogonal complement. 
This abstraction encompasses, for example, metastable manifolds in driven qubits, \textit{degenerate ground-state} subspaces in multi-level atoms~\cite{Fleischhauer2002QuantumMemory}, and dark-state encodings in quantum dots~\cite{Heindel2017DarkExcitonQD}. 
Our goal is to quantify the thermodynamic cost of forcibly steering the system toward the logical-zero manifold within a finite time, and to identify when nonequilibrium initial conditions, specifically those exhibiting the QME, can lower that cost.

The system is initially prepared in a \emph{thermal state} of an \textit{input} Hamiltonian $H_{i}$ at a tunable preparation temperature $T_{\mathrm{prep}}$
\begin{equation}
\rho(0)=\rho_{\mathrm{th}}(T_{\mathrm{prep}})
= \frac{e^{-\beta_{\mathrm{prep}} H_{i}}}{Z_{i}(T_{\mathrm{prep}})}, 
\qquad 
\beta_{\mathrm{prep}} = (k_{B}T_{\mathrm{prep}})^{-1}.
\end{equation}
This preparation temperature plays a central role. By varying it, we generate a continuous family of initial states that differ in their spectral and coherence structure, thereby probing which preparations are thermodynamically most favorable for fast erasure.

At $t=0$, we perform an instantaneous quench, replacing $H_{i}$ with an \textit{erasure Hamiltonian} $H_{f}$ designed such that states within $\mathcal H_0$ are energetically preferred.  
The quench injects work,
\begin{equation}
W_{\mathrm{quench}} = \Tr\!\left[\rho(0)(H_{f}-H_{i})\right],
\end{equation}
which is unavoidable whenever the input and erasure Hamiltonians do not commute.  
This step initiates the erasure protocol by energetically biasing the memory toward the logical zero manifold.

For all $t>0$, the system interacts weakly with a heat bath at temperature $T_b$ and evolves under a completely positive, trace-preserving \textit{Davies-type} Lindblad generator, $\dot\rho(t) = \mathcal{L}[\rho(t)]$, with unique fixed point, the Gibbs state $\rho_{\mathrm{eq}} \propto e^{-\beta_b H_f}$.  
Because this dynamics is Markovian, thermodynamically consistent, and physically implementable, it provides a universal model for irreversible erasure.
The instantaneous population of the target logical state is
\begin{equation}
p_{0}(t) = \Tr[\Pi_0 \rho(t)],
\end{equation}
and erasure is deemed successful once $p_{0}(t)$ exceeds a required fidelity threshold $p_{\mathrm{target}}$ (e.g.\ $p_{\mathrm{target}}=0.95$).
The operational erasure time is therefore $\tau_{\mathrm{erase}} = \inf\{ t : p_{0}(t)\ge p_{\mathrm{target}} \}$.

{\color{black}
The thermodynamic cost of erasure is quantified by the heat released into the
thermal bath. During the dissipative stage, the Hamiltonian $H_f$ is held
fixed, and the heat current into the system is
\begin{subequations}
\begin{equation}
\dot Q_{\rm sys}(t)= \Tr \left[H_f\,\mathcal L(\rho(t))\right].
\end{equation}
The corresponding heat discharged into the bath, which we identify as the
Landauer cost is therefore
\begin{equation}
Q_{\rm bath}(\tau)=-\int_0^\tau \dot Q_{\rm sys}(t)\,dt.
\end{equation}
\end{subequations}
For quasistatic erasure, this quantity approaches the LE, whereas finite-time protocols generally incur additional irreversible contributions. The central question addressed in this work is therefore whether suitably prepared nonequilibrium quantum states can reduce the finite-time erasure cost by suppressing the excitation of the slow relaxation sector of the Liouvillian. As we show below, this mechanism can indeed provide reduced erasure cost within the controlled spectral regime established in our analysis.
}
%-----------------------------------------------------------

\section{Liouvillian spectral analysis and the quantum Mpemba condition}

To uncover the dynamical origin of faster erasure and reduced dissipation, we analyze the spectral structure of the Liouvillian $\mathcal{L}$ acting on the $d^{2}$-dimensional operator space.  
We employ a biorthogonal basis of right and left eigenmodes $\{R_{j}, L_{j}\}$ satisfying
$\mathcal L[R_{j}] = \lambda_{j} R_{j}$, 
$\mathcal L^{\dagger}[L_{j}] = \overline{\lambda_{j}}\,L_{j}$, 
$\Tr[L_{j}^{\dagger} R_{k}] = \delta_{jk}$.
These modes encode all dissipative pathways allowed by the bath and all coherence-decay channels of the system.

The stationary state is the unique zero-mode, $\lambda_{0}=0, R_{0}=\rho_{\mathrm{eq}}$, while all other eigenvalues satisfy $\Re \lambda_{j}<0$ and correspond to exponentially decaying transients. Any deviation from equilibrium admits the expansion $\rho(0)-\rho_{\mathrm{eq}} = \sum_{j\ge1} c_{j} R_{j}$, 
where $c_{j} = \Tr[L_{j}^{\dagger}(\rho(0)-\rho_{\mathrm{eq}})]$,
and the full evolution becomes (see Appendix \ref{Section:Liouvillian} for details)
\begin{equation} \label{rhodynamics}
\rho(t) = \rho_{\mathrm{eq}} + \sum_{j\ge1} c_{j} e^{\lambda_{j} t} R_{j}.
\end{equation}
Physically, the coefficients $c_j$ quantify the degree to which the initial state activates each dissipative relaxation channel.

Ordering the eigenvalues by their decay rates, $\Re \lambda_{1} > \Re \lambda_{2} \ge \dots$, the slowest nonzero mode $j=1$ dominates the relaxation for all but very short times.  Projecting onto the logical-zero population yields
\begin{equation}\label{Eq:population}
\delta p_0 \equiv p_{0}(t)-p_{0}^{\mathrm{eq}}= \sum_{j\ge1} c_{j} e^{\lambda_{j} t} r_{j}^{(0)},  
\qquad 
r_{j}^{(0)} = \Tr[\Pi_{0}R_{j}].
\end{equation}
Rewriting Eq.~\eqref{Eq:population}, we have
\begin{equation}
\delta p_0(t)=c_1r_1^{(0)}e^{\lambda_1t}+\eta(t),
\label{eq:eta_definition}
\end{equation}
where $\eta(t)=\sum_{j\ge2} c_jr_j^{(0)}e^{\lambda_jt}$ contains all higher Liouvillian modes. Unlike the asymptotic limit $t\rightarrow\infty$, the operational erasure time generally occurs at finite times, for which the higher modes need not be negligible. Throughout this work, we therefore require the controlled approximation
\begin{equation}
|\eta(t)|\le \varepsilon |c_1r_1^{(0)}| e^{-\lambda_1t},
\qquad
0<\varepsilon<1,
\label{eq:error_bound}
\end{equation}
over the interval containing the target crossing.
Equation~(\ref{eq:error_bound}) quantifies the accuracy of the single-mode description. 
The target condition $p_0(\tau_{\rm erase})= p_{\rm target}$ becomes
\begin{equation}
p_{\rm target}-p_0^{\rm eq}= c_1r_1^{(0)} e^{\lambda_1\tau_{\rm erase}} +\eta(\tau_{\rm erase}).
\end{equation}
Using the condition (\ref{eq:error_bound}) one can obtain the bound as 
\begin{equation}
\frac1{\lambda_1}\ln \left[\frac{(1-\varepsilon)|c_1r_1^{(0)}|}{|p_{\rm target}-p_0^{\rm eq}|}\right] \le \tau_{\rm erase} \le \frac1{\lambda_1}
\ln \left[\frac{(1+\varepsilon)|c_1r_1^{(0)}|}{|p_{\rm target}-p_0^{\rm eq}|} \right].
\label{eq:bounds_tau}
\end{equation}

Equation~(\ref{eq:bounds_tau}) provides a controlled estimate of the operational erasure time, together with an explicit uncertainty arising from neglected higher relaxation modes. In the limit $\varepsilon\rightarrow 0$, the estimate reduces to the single-mode expression
\begin{equation}
\tau_{\rm erase} \simeq \frac1{\lambda_1}
\ln \left(\frac{|c_1r_1^{(0)}|}{|p_{\rm target}-p_0^{\rm eq}|} \right).
\end{equation}
Therefore, the single-mode approximation should be interpreted as the leading-order description whenever the slow relaxation sector dominates the dynamics over the operational erasure interval.

Thus, within a controlled slow-mode regime and for a common erasure observable, the relevant leading-order condition is
\begin{equation}
(1+\varepsilon_h)|c_1^{(h)}r_1^{(0)}| < (1-\varepsilon_c) |c_1^{(c)}r_1^{(0)}|.
\label{controlled_mpemba_condition}
\end{equation}
implying that a \textit{hotter} preparation $h$ erases \emph{faster} than a \textit{colder} preparation whenever its overlap with the slowest relaxation mode is smaller.   Crucially, we later show that the same condition that accelerates relaxation also reduces the finite-time excess heat above the LB, thereby establishing the QME as a resource for low-dissipation quantum erasure.

Whenever the bound (\ref{eq:error_bound}) is not satisfied, higher Liouvillian modes contribute appreciably, and the operational erasure time must be obtained from the full multimode expansion. The first systematic correction is obtained by retaining the two slowest modes,
\begin{subequations}
\begin{equation}
p_{\rm target}-p_0^{\rm eq}= c_1r_1^{(0)} e^{\lambda_1t} + c_2r_2^{(0)} e^{\lambda_2t},
\end{equation}
which leads to
\begin{equation}
\frac{p_{\rm target}-p_0^{\rm eq}}{c_2r_2^{(0)}e^{\lambda_2t}}= \frac{c_1r_1^{(0)}}
{c_2r_2^{(0)}}e^{-(\lambda_1-\lambda_2)t} +1.
\end{equation}
\end{subequations}
This transcendental equation can be solved numerically and provides the leading correction whenever the operational erasure time approaches the regime where higher relaxation modes remain significant.

\section{Mpemba-enhanced finite-time Landauer erasure}
LP sets the minimum heat dissipation required to erase one bit of information as $\ge k_B T \ln 2$ ($k_B$ is the Boltzmann constant), but this limit is strictly attainable only in the quasistatic regime. Any realistic finite-time erasure necessarily incurs additional dissipation. Understanding \emph{how the structure of nonequilibrium initial states affects this finite-time energetic overhead} is therefore a central challenge in nonequilibrium quantum thermodynamics. In this section, we first illustrate the mechanism for the qutrit model, then extend it to unfold the cause of the reduction in erasure cost in general while suppressing slow Liouvillian modes. 

\begin{figure*}[htpb]
    \centering
    \includegraphics[width=0.95\linewidth]{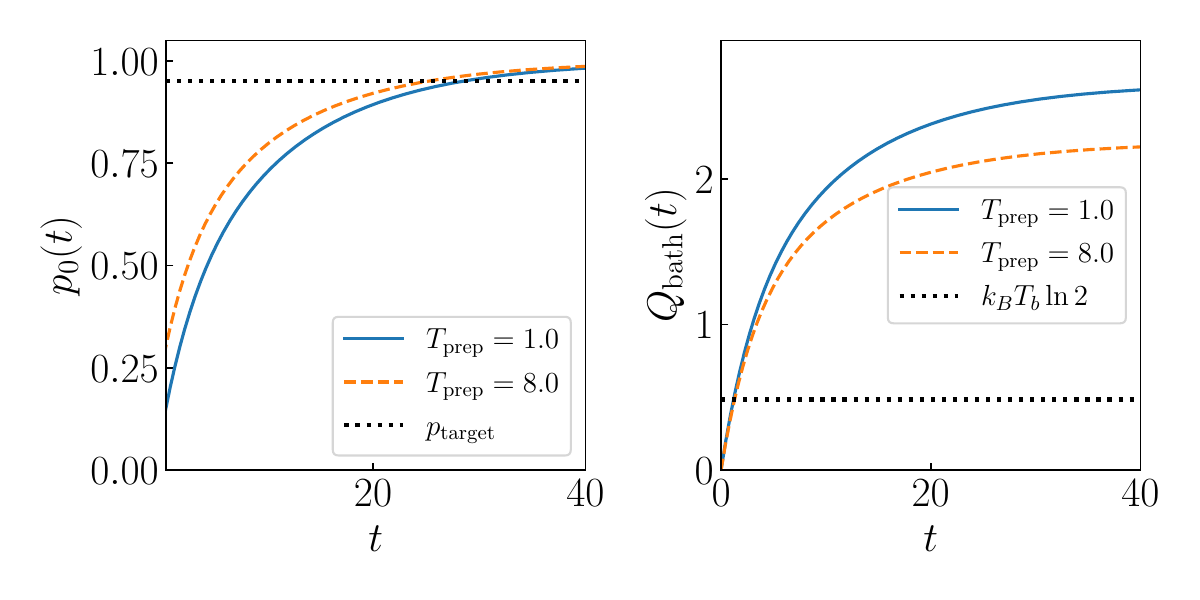}
    \caption{Finite-time Landauer cost and Mpemba advantage in a qutrit memory. Erasure dynamics for a minimal three-level model coupled to a thermal bath at temperature $T_b= 0.6$. The population $p_0(t)$ of the logical-zero level is shown for two thermal preparations at different temperatures $T_{\text{prep}}^{(c)}$ (cold in blue solid line) and $T_{\text{prep}}^{(h)}>T_{\text{prep}}^{(c)}$ (hot in orange dashed line), evolving under the same Davies generator. The hotter preparation reaches the target fidelity $p_{\text{target}}$ at a shorter erasure time $\tau_{\text{erase}}^{(h)}<\tau_{\text{erase}}^{(c)}$, while the accumulated dissipated heat at erasure satisfies $Q_{\text{bath}}^{(h)}(\tau_{\text{erase}}^{(h)}) < Q_{\text{bath}}^{(c)}(\tau_{\text{erase}}^{(c)})$, thereby realizing a genuine quantum Mpemba–enhanced Landauer erasure in which a hotter state both erases faster and dissipates less heat than a colder one. The Hamiltonian parameter values are $\Delta=2.0$, $B=6.0$, $J=0.5$. }
    \label{fig1:plot}
\end{figure*}

\begin{figure*}[htpb]
    \centering
    \includegraphics[width=0.91\linewidth]{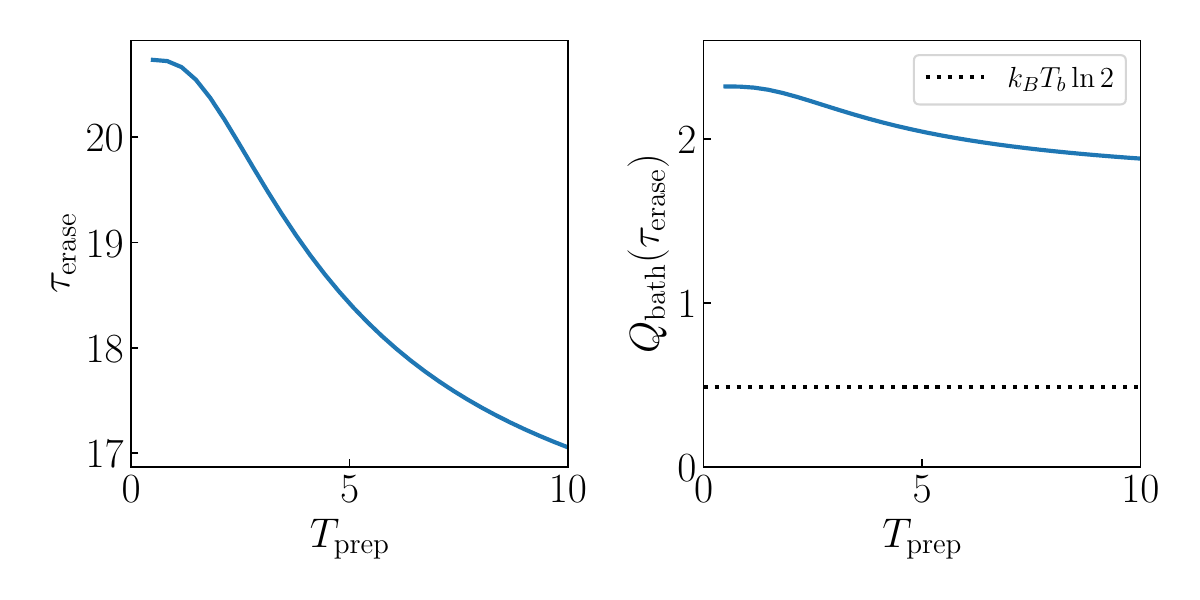}
    \caption{Erasure time and finite-time thermodynamic cost as a function of the preparation temperature in the qutrit memory of Fig.~\ref{fig1:plot}.  Left: operational erasure time $\tau_{\mathrm{erase}}$ versus $T_{\mathrm{prep}}$ for bath temperature $T_b=0.6$ and erasure Hamiltonian.  Right: total dissipated heat at erasure $Q_{\mathrm{bath}}(\tau_{\mathrm{erase}})$, with the dotted line indicating the quasistatic LB $k_{B}T_{b}\ln 2$.  Both quantities show a strong dependence on $T_{\mathrm{prep}}$ and exhibit broad parameter regimes in which hotter preparations simultaneously reduce $\tau_{\mathrm{erase}}$ and $Q_{\mathrm{bath}}$, realizing Mpemba–enhanced Landauer erasure for many hot–cold pairs. The Hamiltonian parameters are the same as Fig.~\ref{fig1:plot}. }
    \label{fig2}
\end{figure*}

\subsection{Qutrit illustration}
To illustrate the proposed mechanism, we consider a minimal three-level
(qutrit) memory whose erasure dynamics are governed by
\begin{equation}
H_f=
\begin{pmatrix}
-B&0&0\\
0&\Delta&J\\
0&J&2\Delta
\end{pmatrix},
\qquad
B>0,\;
\Delta>0,\;
J\ge0.
\end{equation}
The logical state to be erased is encoded in the ground level at energy $-B$, while the excited manifold forms a coherently coupled doublet.  
The off-diagonal matrix element $J$ hybridizes the excited states, ensuring that population flow and coherence decay are intertwined. As a result, the Liouvillian relaxation spectrum acquires nontrivial mode structure. $\Delta$ is the energy scale that sets the bare energies of the two excited levels of the qutrit: one excited state sits at energy $\Delta$ and the other at $2 \Delta$, so their level spacing is $\Delta$. So, $\Delta$ is the characteristic energy (and spacing) of the excited manifold in the final erasure Hamiltonian $H_f$ (see Appendix \ref{Section:qutrit_model} for details).

Diagonalizing $H_f$ yields eigenstates $\{\ket{m}\}$ with energies $\{\varepsilon_m\}$. Given a system--bath coupling operator $V$, the standard Davies construction produces jump operators $A_\omega = \sum_{\varepsilon_m-\varepsilon_n=\omega} \ket{m}\bra{m} V \ket{n}\bra{n}$, with transition rates $\gamma(\omega)$ satisfying detailed balance at temperature $T_b$. These determine the Lindblad generator $\mathcal{L}$ and its spectral data. 

To highlight the role of initial-state structure, we compare erasure starting from two thermal preparations: a cold state $\rho^{(c)}$ at temperature $T_c$ and a hot state $\rho^{(h)}$ at $T_h$, with a fixed bath temperature and erasure Hamiltonian.

Parameter sweeps reveal broad regions where $|c_1^{(h)}| < |c_1^{(c)}|$,
indicating that the hotter preparation has a weaker projection onto the slowest relaxation mode. 
Consequently, it exhibits both a shorter erasure time, $\tau_{\mathrm{erase}}^{(h)} < \tau_{\mathrm{erase}}^{(c)}$, and  a reduced thermodynamic cost $Q_{\mathrm{bath}}^{(h)} < Q_{\mathrm{bath}}^{(c)}$ (Fig. \ref{fig1:plot}). The dependence of the erasure time on $T_{\rm prep}$ and correspondingly the thermodynamic cost is shown in Fig.~\ref{fig2}.

This minimal qutrit model, therefore, provides a transparent demonstration of how nonequilibrium structure, arising from population asymmetries induced by level hybridization, can be harnessed to reduce finite-time LE cost.  It encapsulates the central message of this work: the QME offers a practical pathway to faster, lower-dissipation information erasure beyond the conventional quasistatic LB.

%%%%%%%%%%%%%%%%%%%%%%%%%%%%%%%%%%%%%%%%%%%%%%%%%%%%%%%

\subsection{General finite-time thermodynamics}

The qutrit example illustrates that suppressing the overlap of the initial state with the slowest Liouvillian mode can simultaneously accelerate the approach to the target logical state and reduce the thermodynamic cost of finite-time erasure. We now show that this mechanism is not restricted to the qutrit, but follows directly from the spectral structure of the Davies generator.

During the dissipative stage, the Hamiltonian $H_f$ is fixed, so that the system exchanges only heat with the thermal reservoir. Consequently, the finite-time LE cost is completely determined by the entropy balance. Using Spohn's theorem, the heat released into the bath satisfies
\begin{equation}
Q_{\rm bath}(\tau)= k_B T_b\left[\Sigma(\tau)-\Delta S(\tau)\right],
\label{eq:Spohn_heat}
\end{equation}
where
\begin{subequations}
\begin{equation}
\Sigma(\tau)=\int_0^\tau \sigma(t)\,dt\ge0
\end{equation}
is the integrated entropy production and
\begin{equation}
\Delta S(\tau)=S(\rho(\tau))-S(\rho(0))
\end{equation}
\end{subequations}
is the change in the von Neumann entropy of the system.

The dynamics entering Eq.~\eqref{eq:Spohn_heat} are described by the Liouvillian mode expansion \eqref{Eq:population}. Under the controlled approximation \eqref{eq:error_bound}, the operational erasure time is governed predominantly by the slowest relaxation mode, yielding $\tau_{\rm erase}^{(h)}<\tau_{\rm erase}^{(c)}$
whenever the hotter preparation has the smaller effective overlap with the slow relaxation sector. It remains to determine whether this dynamical speedup also reduces the finite-time entropy production.

To establish this connection, we expand the entropy production around the equilibrium Gibbs state. Since the Davies dynamics drives the system exponentially close to equilibrium, the deviation $\delta\rho(t)=\rho(t)-\rho_{\rm eq}$ admits the Liouvillian decomposition $\delta\rho(t)=\sum_{j\ge1} c_j e^{\lambda_j t}R_j$.
Substituting this expansion into Spohn's entropy production formula and expanding the matrix logarithm around the equilibrium state produces a quadratic form in the Liouvillian amplitudes. The derivation is presented in detail in Appendix~\ref{Section:Mode_entropy}. Retaining the dominant slow-mode contribution gives
\begin{subequations}
\begin{equation}
\Sigma(\tau)=|c_1|^2\frac{1-e^{-2\lambda_1\tau}}{2\lambda_1}\mathcal S_{11}+\mathcal R,
\label{eq:Sigma_quadratic_main}
\end{equation}
where
\begin{equation}
\mathcal S_{11}=-\mathrm{Tr}\!\left[\mathcal L(R_1)\,\mathcal G(R_1)\right] \ge 0
\end{equation}
\end{subequations}
is the quadratic entropy-production coefficient associated with the slowest Liouvillian mode, and $\mathcal R$ collects all remaining multimode contributions together with higher-order nonlinear corrections. Within the controlled regime, the remainder satisfies $|\mathcal R|\ll |c_1|^2\mathcal S_{11}/(2\lambda_1)$, so that the entropy production is governed primarily by the slow-mode overlap. These observations lead to the following general result.

\medskip

\noindent
\textbf{Theorem 1.}
Suppose that the operational erasure dynamics satisfy the controlled slow-mode approximation
\eqref{controlled_mpemba_condition}. Then the hotter preparation reaches the target erasure fidelity sooner than the colder one, i.e.,
\begin{subequations}
\begin{equation}
\tau_{\rm erase}^{(h)}<\tau_{\rm erase}^{(c)}.
\end{equation}
Furthermore, a sufficient condition for the hotter preparation to dissipate less heat is
\begin{equation}
\Delta S^{(h)}-\Delta S^{(c)}> \Sigma^{(h)}(\tau_h)-\Sigma^{(c)}(\tau_c),
\label{eq:heat_condition}
\end{equation}
\end{subequations}
which, by the exact Spohn heat balance \(Q_{\rm bath}=k_BT_b[\Sigma-\Delta S],\)
is equivalent to $Q_{\rm bath}^{(h)}(\tau_h)<Q_{\rm bath}^{(c)}(\tau_c)$.

\medskip

\noindent\textbf{Proof.}
Let two initial preparations be denoted by $\rho^{(h)}(0)$ and $\rho^{(c)}(0)$, and suppose that, following
the instantaneous quench to $H_f$, both states evolve under the same Davies generator $\mathcal L$ at bath temperature $T_b$. Let $\rho_{\rm eq} = \frac{e^{-H_f/T_b}}{\operatorname{Tr}(e^{-H_f/T_b})}$ be the unique stationary Gibbs state, and let $p_0^{(\alpha)}(t)= \operatorname{Tr}\!\left[\Pi_0\rho^{(\alpha)}(t)\right],$ where $\alpha\in\{h,c\}$. A target population $p_{\rm target}$ is fixed, and the operational
erasure time is $\tau_\alpha= \inf\left\{t\geq 0: p_0^{(\alpha)}(t)\geq p_{\rm target} \right\}$.

We assume that: (a) the target crossing occurs in a time interval in which the contribution of the modes $j\geq2$ to the erasure observable is controlled.
(b) For both preparations, $A_\alpha = c_1^{(\alpha)} r_1^{(0)}$ and $d=p_{\rm target}-p_0^{\rm eq}$ have the same sign and $\frac{|A_\alpha|}{|d|}>1$, so that the slow-mode contribution reaches the target value at a finite positive time within the controlled approximation.

% so that the slow sector drives the population monotonically toward the target within the interval considered.

At the target-crossing time, $d= A_\alpha e^{-\lambda_1\tau_\alpha}+
\eta_\alpha(\tau_\alpha)$. Using assumption (i) and the fact that $A_\alpha$ and $d$ have the same sign, we obtain
\begin{equation}
(1-\varepsilon_\alpha)|A_\alpha|e^{-\lambda_1\tau_{\rm erase}^\alpha}
\leq |d|\leq (1+\varepsilon_\alpha) |A_\alpha|e^{-\lambda_1\tau_{\rm erase}^\alpha}.  
\end{equation}
Consequently, we have
\begin{equation}
\frac{1}{\lambda_1}\ln\!\left[
\frac{(1-\varepsilon_\alpha)|A_\alpha|}{|d|}\right] \leq \tau_{\rm erase}^\alpha
\leq
\frac{1}{\lambda_1}\ln\!\left[ \frac{(1+\varepsilon_\alpha)|A_\alpha|} {|d|}\right].   
\end{equation}
which is equivalent to $(1+\varepsilon_h)|A_h|<(1-\varepsilon_c)|A_c|$. Therefore, $\tau_h<\tau_c$.

We now consider the thermodynamic comparison. For a Davies generator, the Spohn entropy-production rate is
\begin{subequations}
\begin{equation}\label{Eq:sigma_alpha}
\sigma^{(\alpha)}(t) = -\operatorname{Tr}\!\left[ \mathcal L(\rho^{(\alpha)}(t)) \left(\ln\rho^{(\alpha)}(t)-\ln\rho_{\rm eq} \right) \right] \geq0. 
\end{equation}
Using $\ln\rho_{\rm eq}= -\frac{H_f}{T_b}
-\ln Z_b$, together with trace preservation, we get
\begin{equation}
\sigma^{(\alpha)}(t)= \frac{d}{dt}S(\rho^{(\alpha)}(t)) -\frac{\dot Q_{\rm sys}^{(\alpha)}(t)}{T_b},
\end{equation}
Integration from $0$ to $\tau_\alpha$ yields
\begin{equation}
\Sigma^{(\alpha)}(\tau_\alpha):= \int_0^{\tau_\alpha} \sigma^{(\alpha)}(t)\,dt= \Delta S^{(\alpha)} - \frac{Q_{\rm sys}^{(\alpha)}(\tau_\alpha)}{T_b}.    
\end{equation}
\end{subequations}
The exact heat balance is
\begin{subequations}
\begin{equation}
Q_{\rm bath}^{(\alpha)}(\tau_\alpha)=T_b \left[ \Sigma^{(\alpha)}(\tau_\alpha)
- \Delta S^{(\alpha)} \right].
\label{eq:exact_bath_heat}
\end{equation}
This identity contains both the change in the system entropy and the total entropy production. Since the Hamiltonian is held fixed throughout the dissipative relaxation stage, no work is performed after the initial quench. The first law, therefore, reduces to pure heat exchange.
For completeness, the same result follows from the first law:
\begin{equation}
 Q_{\rm bath}^{(\alpha)}(\tau_\alpha)
=-\Delta E^{(\alpha)}.   
\end{equation}
\end{subequations}
Thus, during autonomous Davies relaxation at fixed $H_f$, the free-energy decrease and the integrated entropy production are not independent
contributions.

Near equilibrium, let
\begin{equation}
\delta\rho^{(\alpha)}(t) = \rho^{(\alpha)}(t)-\rho_{\rm eq}
=\sum_{j\geq1}c_j^{(\alpha)}e^{\lambda_jt}R_j. 
\end{equation}
Expanding the logarithm around $\rho_{\rm eq}$ in Eq.~\eqref{Eq:sigma_alpha} gives
\begin{equation}
\sigma^{(\alpha)}(t)= -\operatorname{Tr}\!\left[\mathcal L(\delta\rho^{(\alpha)}(t))
\mathcal G(\delta\rho^{(\alpha)}(t))
\right]+ O\!\left(\|\delta\rho^{(\alpha)}(t)\|^3\right).    
\end{equation}
Hence, after some algebraic calculation (see Appendix \ref{Section:Mode_entropy} for details), we have
\begin{equation}
\Sigma^{(\alpha)}(\tau_\alpha)=\sum_{j,k\geq1}
c_j^{(\alpha)} \big(c_k^{(\alpha)}\big)^*
\frac{e^{(\lambda_j+\overline{\lambda_k})\tau_\alpha}-1}{\lambda_j+\overline{\lambda_k}}
\mathcal S_{jk}+O\!\left(\|\delta\rho^{(\alpha)}\|^3\right),    
\end{equation}
where
\begin{equation}
 \mathcal S_{jk}= -\operatorname{Tr}\!\left[ \mathcal L[R_j]\,\mathcal G[R_k]\right].   
\end{equation}
Separating the $j=k=1$ contribution from all remaining terms gives
\begin{equation}
\Sigma^{(\alpha)}(\tau_\alpha)= |c_1^{(\alpha)}|^2 \frac{1-e^{-2\lambda_1\tau_\alpha}} {2\lambda_1}
\mathcal S_{11}+ \mathcal R_\alpha,   
\end{equation}
with $|\mathcal R_\alpha|\leq\delta_\alpha$.
The constants $\delta_\alpha$ quantify the cumulative contribution of
the neglected higher Liouvillian modes and nonlinear corrections over
the operational erasure interval. In practice they can be estimated
numerically from the full Liouvillian spectrum or bounded analytically
whenever a spectral gap separates the slowest relaxation mode from the
remaining modes.

Subtracting Eq.~\eqref{eq:exact_bath_heat} for the two preparations,
we find
\begin{align}
Q_{\rm bath}^{(h)}(\tau_h)-Q_{\rm bath}^{(c)}(\tau_c)= T_b \Big[\Sigma^{(h)}(\tau_h)-\Sigma^{(c)}(\tau_c) -\big(\Delta S^{(h)}-\Delta S^{(c)} \big)\Big].
\label{eq:bath_heat_difference}
\end{align}
Using the quadratic expansions, we get
\begin{align}
\Sigma^{(h)}(\tau_h)-\Sigma^{(c)}(\tau_c) =&\;\frac{\mathcal S_{11}}{2\lambda_1}\Big[|c_1^{(h)}|^2 \big(1-e^{-2\lambda_1\tau_h}\big) \nonumber\\
&\hspace{4.7em}-|c_1^{(c)}|^2\big(1-e^{-2\lambda_1\tau_c}\big)\Big]
+\mathcal R_h-\mathcal R_c.
\end{align}
The remainder satisfies $\mathcal R_h-\mathcal R_c \leq |\mathcal R_h|+|\mathcal R_c| \leq \delta_h+\delta_c$. Therefore, Eq.~\eqref{eq:bath_heat_difference} evolves to
\begin{align}
Q_{\rm bath}^{(h)}(\tau_h)-Q_{\rm bath}^{(c)}(\tau_c)\leq T_b\Bigg\{
&\frac{\mathcal S_{11}}{2\lambda_1}
\Big[|c_1^{(h)}|^2 \big(1-e^{-2\lambda_1\tau_h}\big) \nonumber\\
&\hspace{3.5em}
-|c_1^{(c)}|^2 \big(1-e^{-2\lambda_1\tau_c}\big)\Big]\nonumber\\
&+\delta_h+\delta_c- \big(\Delta S^{(h)}-\Delta S^{(c)} \big)\Bigg\}.
\end{align}
Condition \eqref{eq:SI5_explicit_sufficient_heat_condition} makes the right-hand side strictly negative. It therefore follows that $Q_{\rm bath}^{(h)}(\tau_h) < Q_{\rm bath}^{(c)}(\tau_c)$.

\hfill$\square$

\medskip

The theorem, therefore, provides two independent results. The controlled slow-mode condition guarantees a dynamical Mpemba speedup provided the higher-mode remainder remains bounded over the operational erasure interval. A reduction of the absolute LE cost is a distinct thermodynamic statement and follows only after combining the exact Spohn entropy balance with the additional entropy condition Eq.~\eqref{eq:SI5_explicit_sufficient_heat_condition}. Consequently, suppression of the slow Liouvillian mode alone should not be interpreted as implying lower heat dissipation. Finally, when the operational erasure time lies outside the controlled slow-mode regime, the complete multimode Liouvillian expansion must be used instead of the leading-order approximation.

%-------------------------------------------------------------

%---------------------------------------------------------------

%-------------------------------------------------------------

\section{Physical implementation and feasibility}
The proposed Mpemba-enhanced finite-time LE protocol is experimentally feasible using several existing quantum technologies that simultaneously provide state preparation, engineered dissipation, and high-fidelity readout. The mentioned features have been demonstrated in trapped ions~\cite{Barreiro2011OpenSystem,Lin2013Dissipative,Zhang2025StrongMpemba}, superconducting qubits~\cite{Shankar2013Autonomous,Chen2025HardwareEfficient}, and NV-center systems~\cite{Liu2019NVControl,Kurokawa2024NV0Control}.

Superconducting quantum circuits constitute perhaps the most direct implementation platform of the present protocol. 
Coupling the qubit to a microwave resonator or a tunable dissipative environment generates controlled Lindblad dynamics whose relaxation rates can be engineered over several orders of magnitude. Importantly, superconducting circuits have already provided experimental demonstrations of finite-time LE through calorimetric measurements of dissipated heat, while dispersive readout enables continuous monitoring of excited-state populations with high fidelity~\cite{saira2020nonequilibrium,PhysRevLett.120.210601}. Since our protocol requires comparing the erasure dynamics of different initial preparations evolving under the same dissipative generator, these systems offer an ideal platform for directly measuring both the reduction in erasure time and the corresponding decrease in finite-time heat dissipation predicted by our theory.

Trapped-ion platforms provide an equally attractive realization because they combine exceptionally long coherence times with programmable open-system dynamics~\cite{An2015,Huber2008}.
Recent experiments have demonstrated the QME in trapped ions~\cite{PhysRevLett.133.010402,PhysRevLett.133.010403} by engineering dissipation that produces an inversion of relaxation speeds between different initial states, thereby directly verifying the Liouvillian mechanism underlying our theory. The same experimental platform can be employed to realize quantum thermodynamic protocols, including measurements of work distributions and nonequilibrium entropy production.

Semiconductor quantum dots coupled to electronic reservoirs provide another natural setting for testing our predictions. 
Recent theoretical and experimental studies have shown that quantum-dot systems exhibit Mpemba-type relaxation anomalies~\cite{PhysRevLett.131.080402} arising from the spectral properties of the corresponding Liouvillian. Because charge occupation probabilities can be monitored continuously using nearby quantum point contacts or single-electron transistors, both the operational erasure time and the relaxation dynamics can be determined with high temporal resolution. Moreover, calorimetry provides a direct route to measuring the finite-time heat exchanged with the reservoirs, enabling a quantitative verification of the modified LB derived in this work.

Solid-state spin platforms, like NV centers~\cite{Neumann2013NVthermo,chatterjee2025direct,schnepper2025experimental}, offer complementary advantages arising from their precise coherent control and exceptionally accurate state tomography. Recent experiments~\cite{schnepper2025experimental,chatterjee2025direct} have directly observed QME in this platform, which routinely performs complete quantum process tomography, allowing reconstruction of the Liouvillian spectrum and the overlap of the initial state with its slow relaxation modes. These capabilities make them well-suited for testing the erasure cost.

%--------------------------------------------------

\section{Conclusion and Outlook}
To summarize, we have demonstrated that nonequilibrium quantum initial states can be harnessed as a \textit{genuine thermodynamic resource} to lower the finite-time cost of LE. By combining a Liouvillian-mode decomposition with a refined finite-time LE, we showed that QME, in which hotter states have a strongly reduced overlap with the slowest relaxation mode, can both speed up erasure and reduce the dissipated heat compared to more \textit{conventional} colder preparations, all while remaining fully consistent with the quasistatic LB. A minimal multi-level model illustrates this mechanism explicitly and reveals broad parameter regimes in which Mpemba-enhanced erasure arises from the interplay of population asymmetries, coherences, and Liouvillian non-normality, thereby linking dynamical relaxation anomalies directly to practical gains in information erasure. We emphasize, however, that the qutrit serves only as an illustrative example; the theoretical framework itself is formulated at the level of the Liouvillian spectral decomposition and is not tied to any particular finite-dimensional realization.

Our findings place the QME in a new light: not merely as a relaxation phenomenon, but as a resource for finite-time thermodynamics and information processing. It highlights that the thermodynamic cost of erasure is not determined solely by the target state and bath temperature, but also by how cleverly one engineers the initial nonequilibrium state and its decomposition into dynamical modes. This perspective suggests a broader design principle for thermal operations in quantum technologies, where mode-structure engineering becomes as important as controlling energies and entropies.

Looking ahead, several theoretical extensions are natural. 
An important open question in this direction is to determine the extent to which the present mechanism survives in systems with unbounded spectra, where additional physical effects absent in finite-dimensional models may influence the relaxation dynamics and thermodynamic cost of erasure. Extending the present framework beyond finite-dimensional Davies dynamics to include non-Markovian reservoirs, strong system-bath coupling, and driven or periodically modulated environments would clarify the robustness and generality of Mpemba-enhanced erasure as a finite-time thermodynamic resource.

Generalizing from single-bit or few-level memories to many-body registers and correlated baths could reveal collective Mpemba mechanisms and scaling laws for energy-efficient quantum memories. It would also be interesting to connect our Liouvillian-mode picture with concepts such as thermodynamic length~\cite{Crooks2007ThermodynamicLength}, quantum speed limits~\cite{Deffner2017QSL}, sensing~\cite{gsh7r7ms,DePasquale2016,PritamQST2,Potts2019} and optimal control~\cite{Glaser2015QOC}, in order to identify provably optimal erasure protocols that exploit Mpemba-type mode suppression.

On the experimental side, our control strategies based on temperature tuning, coherence engineering, and Hamiltonian shaping of Liouvillian spectra are compatible with state-of-the-art platforms such as superconducting circuits, trapped ions, semiconductor quantum dots, and solid-state spins, where both high-fidelity state preparation and relaxation are routinely available. Systematic tests of quantum Mpemba–enhanced LE in these architectures, for instance via calorimetric heat measurements or trajectory-resolved monitoring of the relaxation dynamics, would provide direct benchmarks of our predictions. In the longer term, embedding Mpemba-optimized erasure cycles into functional devices like quantum memories~\cite{Simon2010QuantumMemories,bose,chiara}, logic elements~\cite{Ren2011ReversibleSFQ}, thermal machine~\cite{chattopadhyay2019relativistic,vinjanampathy2016quantum,chattopadhyay2021quantum,scully_science,scully_pnas,chattopadhyay2020non} or autonomous Maxwell demons~\cite{maxwell1871theory,maruyama2009colloquium,NajeraSantos2020AutonomousDemon}, could establish nonequilibrium initial-state design as a standard tool for minimizing energy consumption in nanoscale information processing and for exploring the ultimate thermodynamic limits of quantum technologies.

%\bibliography{ref}

%merlin.mbs apsrev4-1.bst 2010-07-25 4.21a (PWD, AO, DPC) hacked
%Control: key (0)
%Control: author (0) dotless jnrlst
%Control: editor formatted (1) identically to author
%Control: production of article title (0) allowed
%Control: page (1) range
%Control: year (0) verbatim
%Control: production of eprint (0) enabled
%

\newpage
\onecolumngrid
\appendix
\section{ Liouvillian spectral theory and biorthonormal modes \label{Section:Liouvillian}}

We consider a finite-dimensional Hilbert space $\mathcal H$ of dimension $d$ and a Lindblad generator 
\begin{equation}
\dot{\rho}(t)=\mathcal L[\rho(t)],
\end{equation}
in the Schr\"odinger picture.  Throughout, we assume that $\mathcal L$ is a Davies generator obtained in the weak-coupling, Markovian, and secular limits \cite{breuer_petruccione,davies1974,spohn1978}, describing thermalization with a reservoir at inverse temperature $\beta_b=(k_B T_b)^{-1}$.  

Under standard assumptions, namely primitivity of the semigroup and the absence of symmetry-protected stationary subspaces, the channel $e^{t\mathcal L}$ is completely positive, trace-preserving (CPTP) and relaxes to a unique, full-rank Gibbs state,
\begin{equation}
\rho_{\mathrm{eq}}
= \frac{e^{-\beta_b H_f}}{\mathrm{Tr}\!\left[e^{-\beta_b H_f}\right]} .
\end{equation}

We endow the operator space $\mathcal B(\mathcal H)$ with the Hilbert--Schmidt (HS) inner product
\begin{equation}
\langle A,B\rangle_{\mathrm{HS}} = \mathrm{Tr}[A^\dagger B].
\end{equation}
With this inner product, the generator $\mathcal L$ is generally non-normal ($\mathcal L^\dagger \mathcal L \neq \mathcal L\mathcal L^\dagger$), and therefore right and left eigenmodes must be treated separately. Let the spectrum of $\mathcal L$ consist of eigenvalues 
\begin{equation}
\mathrm{spec}(\mathcal L)=\{\lambda_j\}_{j=0}^{d^2-1}, \qquad \Re\lambda_j\le 0,
\end{equation}
with a unique zero eigenvalue $\lambda_0=0$ whose right eigenmode is the equilibrium state $R_0=\rho_{\mathrm{eq}}$.

Right and left eigenmodes are defined by
\begin{subequations}
\begin{align}
\mathcal L[R_j] &= \lambda_j R_j, \\
\mathcal L^\dagger[L_j] &= \overline{\lambda_j}\, L_j ,
\end{align}
\end{subequations}
where $\mathcal L^\dagger$ is the HS-adjoint.  
We impose biorthonormality,
\begin{equation}
\mathrm{Tr}\!\left[L_j^\dagger R_k\right] = \delta_{jk}.
\end{equation}

Davies generators are generically diagonalizable %(semisimple) 
on $\mathcal B(\mathcal H)$, except for parameter choices where Jordan blocks may appear. Under the assumption of diagonalizability, the set $\{R_j\}$ forms a basis of an operator space.  
Thus, any operator $X$ admits the decomposition
\begin{equation}
X=\sum_{j=0}^{d^2-1} \mathrm{Tr}[L_j^\dagger X]\, R_j .
\end{equation}

If the generator is not diagonalizable, generalized eigenvectors must be introduced. The corresponding time dependence contains terms of the form $t^m e^{\lambda_jt}$, rather than pure exponentials. In that case, all statements below remain applicable only after replacing individual eigenmodes by the corresponding generalized eigenspaces and explicitly retaining the polynomial prefactors.

Describing the initial deviation from equilibrium as
\begin{equation}
\rho(0)-\rho_{\mathrm{eq}}=\sum_{j\ge 1} c_j R_j,
\qquad
c_j = \mathrm{Tr}[L_j^\dagger(\rho(0)-\rho_{\mathrm{eq}})],
\end{equation}
the full-time evolution becomes
\begin{equation}
\rho(t) = \rho_{\mathrm{eq}} + \sum_{j\ge 1} c_j e^{\lambda_j t} R_j,
\end{equation}
with all nonstationary modes decaying due to $\Re\lambda_j<0$ for $j\ge 1$.

As $\mathcal L$ is not normal, its left and right eigenmodes differ in general.  
Observable relaxation properties, including sensitivity to initial state, anomalously fast cooling, and Mpemba inversions, are governed by these non-orthogonal modes.  
This non-normality is the key mathematical structure enabling Mpemba-type acceleration in quantum dissipative systems.
The Liouvillian overlap coefficients
\begin{equation}
c_j = \mathrm{Tr}\!\big[ L_j^\dagger (\rho(0)-\rho_{\mathrm{eq}} ) \big]
\end{equation}
quantify how the initial state decomposes into dynamical relaxation modes.

The contribution of each Liouvillian mode to the dynamics is observable-dependent. For
an observable $O$, one obtains
\begin{equation}
\langle O\rangle_t-\langle O\rangle_{\rm eq} = \sum_{j\geq1}  c_j e^{\lambda_jt}o_j,
\qquad
 o_j:= \Tr(OR_j).
    \label{eq:SI1_observable_expansion}
\end{equation}
A mode with a large coefficient $c_j$ does not affect the relaxation of
$O$ if $o_j=0$.

For the erasure observable $O=\Pi_0$, define
\begin{equation}
r_j^{(0)} :=\Tr(\Pi_0R_j).
\end{equation}
Then
\begin{equation}
 p_0(t)-p_0^{\rm eq} = \sum_{j\geq1}
    c_jr_j^{(0)}e^{\lambda_jt}. \label{eq:SI1_erasure_observable_expansion}
\end{equation}
Thus, the erasure-relevant modal amplitude is
$c_jr_j^{(0)}$, rather than $c_j$ alone.
Equation~\eqref{eq:SI1_erasure_observable_expansion} may be separated as
\begin{subequations}
\begin{equation}
p_0(t)-p_0^{\rm eq} =
c_1r_1^{(0)}e^{\lambda_1t}+\eta_{\Pi_0}(t),
    \label{eq:SI1_slow_plus_remainder}
\end{equation}
where
\begin{equation}
\eta_{\Pi_0}(t):= \sum_{j\geq2}  c_jr_j^{(0)}e^{\lambda_jt}.
\end{equation}
\end{subequations}

The single-mode approximation is controlled only over an interval in
which
\begin{equation}
 |\eta_{\Pi_0}(t)| \leq \varepsilon |c_1r_1^{(0)}|e^{-\lambda_1t},
\qquad
0\leq\varepsilon \ll 1.    \label{eq:SI1_observable_remainder_bound}
\end{equation}
The bound \eqref{eq:SI1_observable_remainder_bound} generally does not hold at very short times because the higher Liouvillian modes have not yet decayed appreciably. Since $e^{\lambda_j t} \approx 1$ for all $j$ when $t\rightarrow 0$, the remainder $\eta_{\Pi_0}(t)$ may be comparable to the slow-mode contribution. The approximation becomes controlled only after the faster modes are suppressed relative to the slowest mode, or equivalently when the spectral-gap-induced decay of the higher modes renders $\varepsilon << 1$.
In particular, when the operational erasure time is comparable to or shorter than the decay times of several higher modes, all corresponding terms in Eq.~\eqref{eq:SI1_erasure_observable_expansion} must be retained.

Within a controlled slow-mode regime and for a common erasure observable, the
relevant leading-order condition is
\begin{equation}
(1+\varepsilon_h)|c_1^{(h)}r_1^{(0)}|
<(1-\varepsilon_c) |c_1^{(c)}r_1^{(0)}|.
\label{eq:SI1_controlled_mpemba_condition}
\end{equation}
If condition \eqref{eq:SI1_controlled_mpemba_condition} is violated, the contribution of the higher Liouvillian modes is no longer uniformly small compared with the slowest mode over the operational erasure interval. Consequently, the target-crossing time cannot be inferred from the leading coefficient $c_1 r_1^{(0)}$ alone. Instead, the erasure time must be obtained from the full multimode expression.

In the case of a nondegenerate Hamiltonian, the dynamics commonly separates into invariant sectors associated with populations and coherences in the energy eigenbasis. The diagonal sector obeys a Pauli master equation,
\begin{equation}
\dot{\mathbf p}(t)= W\mathbf p(t),
\end{equation}
where the off-diagonal entries of $W$ are generated by the
nonzero-Bohr-frequency components of the system--bath coupling.
Writing the right and left eigenvectors of $W$ as
$\mathbf v_j$ and $\mathbf w_j$, respectively, one obtains
\begin{equation}
c_j=\mathbf w_j^\top \bigl( \mathbf p(0)-\mathbf p_{\rm eq}\bigr)
\end{equation}
for population modes. The condition $[V,H_f]=0$ implies that $V$ contains only a zero-Bohr-frequency component and therefore generally produces dephasing without energy
exchange. For a general initial state,
\begin{equation}
\rho(0)=\sum_{m,n}\rho_{mn}^{(0)} |m\rangle\langle n|,
\end{equation}
the spectral coefficient is
\begin{equation}
c_j=\sum_{m,n} \rho_{mn}^{(0)} \langle n|L_j^\dagger|m\rangle -\Tr(L_j^\dagger\rho_{\rm eq}).
    \label{eq:SI1_coherence_coefficients}
\end{equation}
Off-diagonal matrix elements of $\rho(0)$ contribute to $c_j$ only when the corresponding left eigenmode $L_j$ has support in the coherence sector. Moreover, a coherence contribution to $c_j$ affects the erased population only if the associated right mode has
\begin{equation}
r_j^{(0)}=\Tr(\Pi_0R_j)\neq0.
\end{equation}
For a nondegenerate Davies generator with a diagonal erasure projector, pure coherence modes commonly satisfy $r_j^{(0)}=0$ and do not directly modify $p_0(t)$. Coherence-assisted population erasure therefore requires a mechanism that couples the population and coherence sectors of the Liouvillian, such as degenerate energy levels or Bohr frequencies, engineered dissipative couplings, or a non-diagonal logical projector.

\section{Qutrit model \label{Section:qutrit_model}}

During the dissipative erasure stage, the qutrit is governed by 
\begin{equation}
H_f=
\begin{pmatrix}
-B & 0 & 0\\
0 & \Delta & J\\
0 & J & 2\Delta
\end{pmatrix},
\qquad B>0,\ \Delta>0,\ J\ge 0,
\label{eq:SI_Hf_matrix}
\end{equation}
in the computational basis $\{|0\rangle,|1\rangle,|2\rangle\}$. The excited manifold
$\{|1\rangle,|2\rangle\}$ is diagonalized by an angle $\theta$ defined by
\begin{equation}
\tan(2\theta)=\frac{2J}{\Delta},\qquad \theta\in(0,\pi/2)\ (J>0).
\label{eq:SI_theta_def}
\end{equation}
The eigenenergies are
\begin{equation}
\varepsilon_0=-B,\qquad
\varepsilon_{\pm}=\frac{3\Delta\pm \Omega}{2},\qquad
\Omega=\sqrt{\Delta^2+4J^2},
\label{eq:SI_eigs}
\end{equation}
with eigenstates
\begin{equation}
|+\rangle=\cos\theta\,|1\rangle+\sin\theta\,|2\rangle,\qquad
|-\rangle=-\sin\theta\,|1\rangle+\cos\theta\,|2\rangle.
\label{eq:SI_eigvecs}
\end{equation}
We denote projectors $\Pi_0=|0\rangle\langle0|$ and $\Pi_{\pm}=|\pm\rangle\langle\pm|$.
The (potentially relevant) Bohr frequencies are
\begin{align}
\omega_{+0}&=\varepsilon_+-\varepsilon_0=\varepsilon_+ + B, \nonumber\\
\omega_{-0}&=\varepsilon_- -\varepsilon_0=\varepsilon_- + B, \nonumber\\
\omega_{+-}&=\varepsilon_+-\varepsilon_-=\Omega .
\label{eq:SI_bohr}
\end{align}

\subsection*{A. System--bath coupling operator}
The weak-coupling interaction is taken as $H_{\mathrm{int}}=V\otimes B_{\mathrm{bath}}$ with a
Hermitian system operator $V$. A minimal \textit{reset} coupling that connects the logical level to the excited manifold is
\begin{equation}
V_{\mathrm{reset}}
=g\Big(|0\rangle\langle1|+|1\rangle\langle0|+|0\rangle\langle2|+|2\rangle\langle0|\Big),
\label{eq:SI_V_reset}
\end{equation}
where $g$ sets the overall relaxation timescale (equivalently absorbed into the rates below).
If additional scans include other noise channels, they can be specified, e.g.
\begin{subequations}
\begin{align}
V_{\mathrm{exch}}&=g_e\left(|1\rangle\langle2|+|2\rangle\langle1|\right), \label{eq:SI_V_exch}\\
V_{\phi}&=g_\phi\left(|1\rangle\langle1|-|2\rangle\langle2|\right). \label{eq:SI_V_deph}
\end{align}
\end{subequations}

\subsection*{B. Jump operators $A_\omega$ and Rates $\gamma(\omega)$}
Let $H_f=\sum_m \varepsilon_m \Pi_m$ be the spectral decomposition in the energy basis
$\{|0\rangle,|-\rangle,|+\rangle\}$. The Davies (secular) decomposition defines the jump operators
\begin{equation}
A_\omega=\sum_{\varepsilon_m-\varepsilon_n=\omega}\Pi_m\,V\,\Pi_n,
\qquad
A_{-\omega}=A_\omega^\dagger\ \ (V=V^\dagger).
\label{eq:SI_Aomega_general}
\end{equation}
For $V_{\mathrm{reset}}$ in Eq.~\eqref{eq:SI_V_reset} only ground--excited transitions appear, giving
\begin{align}
A_{\omega_{+0}}&=\Pi_+ V_{\mathrm{reset}}\Pi_0
=g(\cos\theta+\sin\theta)\,|+\rangle\langle0|, \label{eq:SI_A_plus0}\\
A_{\omega_{-0}}&=\Pi_- V_{\mathrm{reset}}\Pi_0
=g(\cos\theta-\sin\theta)\,|-\rangle\langle0|. \label{eq:SI_A_minus0}
\end{align}
If $V_{\mathrm{exch}}$ is included, then an additional transition within the excited manifold arises,
$A_{\omega_{+-}}=\Pi_+ V_{\mathrm{exch}}\Pi_- \propto |+\rangle\langle-|$.
If $V_{\phi}$ is included, then an $\omega=0$ component $A_{0}=\sum_m \Pi_m V_{\phi}\Pi_m$ produces
pure dephasing.

We set $\hbar=k_B=1$ and denote the bath inverse temperature by $\beta_b=1/T_b$.
The Davies generator is fully determined by the rates $\gamma(\omega)$ satisfying the KMS
(detailed balance) condition
\begin{equation}
\gamma(-\omega)=e^{-\beta_b\omega}\,\gamma(\omega)\qquad (\omega>0),
\label{eq:SI_detailed_balance}
\end{equation}
ensuring that $\rho_{\mathrm{eq}}\propto e^{-\beta_b H_f}$ is stationary.
A standard bosonic-bath choice is
\begin{equation}
\gamma(\omega)=2\pi\,\mathcal{J}(|\omega|)
\begin{cases}
n_b(|\omega|)+1, & \omega>0,\\[2pt]
n_b(|\omega|), & \omega<0,
\end{cases}
\qquad
n_b(\nu)=\frac{1}{e^{\beta_b \nu}-1},
\label{eq:SI_gamma_boson}
\end{equation}
with spectral density $\mathcal{J}(\nu)$ (for $\nu\ge 0$), e.g. Ohmic with cutoff
\begin{equation}
\mathcal{J}(\nu)=\kappa\,\nu\,e^{-\nu/\omega_c},
\label{eq:SI_ohmic}
\end{equation}
where $\kappa$ controls the overall dissipative timescale and $\omega_c$ is the cutoff.

\subsection*{C. Lindblad generator}
The Schr\"odinger-picture evolution is
\begin{equation}
\dot\rho(t)=\mathcal{L}[\rho(t)],
\label{eq:SI_master_equation}
\end{equation}
with the Davies (GKLS) generator
\begin{equation}
\mathcal{L}[\rho]
=-i[H_f+H_{\mathrm{LS}},\rho]
+\sum_{\omega}\gamma(\omega)
\left(
A_\omega\rho A_\omega^\dagger
-\frac12\{A_\omega^\dagger A_\omega,\rho\}
\right).
\label{eq:SI_Davies_generator}
\end{equation}
The sum runs over all Bohr frequencies $\omega$ for which $A_\omega\neq 0$.

\section{Mode-Resolved Entropy Production and Conditional Heat Reduction \label{Section:Mode_entropy}}

The heat current into the system is defined by
\begin{subequations}
\begin{equation}
\dot Q_{\rm sys}(t)= \Tr\!\left[H_f\mathcal L(\rho(t))\right].
\end{equation}
Because $H_f$ is time independent during the relaxation stage,
\begin{equation}
Q_{\rm sys}(\tau)=\int_0^\tau\dot Q_{\rm sys}(t)\,dt= E(\tau)-E(0)\equiv\Delta E(\tau).
\label{eq:SI5_Qsys_first_law}
\end{equation}
\end{subequations}
The heat discharged into the bath is therefore
\begin{equation}
Q_{\rm bath}(\tau):= -Q_{\rm sys}(\tau)
=E(0)-E(\tau).
\label{eq:SI5_Qbath_definition}
\end{equation}
With this convention, $Q_{\rm bath}>0$ denotes a positive amount of energy transferred from the system to the environment and is the quantity identified as the Landauer erasure cost.

Let
\begin{equation}
S(\rho)=-\Tr(\rho\ln\rho),
\end{equation}
denote the von Neumann entropy. The Spohn entropy-production rate is
\begin{equation}
\dot\Sigma(t)= -\Tr\!\left[ \mathcal L(\rho(t))\bigl( \ln\rho(t)-\ln\rho_{\rm eq}\bigr) \right] \geq0.
\label{eq:SI5_Spohn_rate}
\end{equation}
Using
\begin{equation}
\ln\rho_{\rm eq}=-\beta_bH_f-\ln Z_b
\end{equation}
and trace preservation, Eq.~\eqref{eq:SI5_Spohn_rate} becomes
\begin{subequations}
\begin{equation}
\dot\Sigma(t)=\frac{d}{dt} S(\rho(t))
-\beta_b\dot Q_{\rm sys}(t).
\label{eq:SI5_entropy_balance_rate}
\end{equation}
Integration gives the exact entropy balance
\begin{equation}
\Sigma(\tau):= \int_0^\tau\dot\Sigma(t)\,dt
=\Delta S(\tau)-\beta_bQ_{\rm sys}(\tau)
\geq0,
\label{eq:SI5_entropy_balance_integrated}
\end{equation}
where
\begin{equation}
\Delta S(\tau)=S(\rho(\tau))-S(\rho(0)).
\end{equation}
\end{subequations}
Equivalently, the heat discharged into the bath is
\begin{equation}
Q_{\rm bath}(\tau)= k_BT_b
\left[\Sigma(\tau)-\Delta S(\tau)
\right].
\label{eq:SI5_exact_bath_heat}
\end{equation}

For deviations sufficiently close to $\rho_{\rm eq}$, the matrix
logarithm admits the expansion
\begin{equation}
\ln(\rho_{\rm eq}+\delta\rho)=\ln\rho_{\rm eq}
+\mathcal G[\delta\rho] + O(\|\delta\rho\|^2),
\end{equation}
where $\mathcal G$ is the Fr\'echet derivative of the logarithm at
$\rho_{\rm eq}$. Substitution into Eq.~\eqref{eq:SI5_Spohn_rate} gives
\begin{equation}
\dot\Sigma(t)= -\Tr\!\left[\mathcal L[\delta\rho(t)] \mathcal G[\delta\rho(t)]\right] + O(\|\delta\rho(t)\|^3).
\label{eq:SI5_sigma_quadratic}
\end{equation}
The quadratic contribution is
\begin{equation}
\dot\Sigma(t)= \sum_{j,k\geq1}
c_jc_k^* e^{(\lambda_j+\overline{\lambda_k})t}
\mathcal S_{jk} + O(\|\delta\rho(t)\|^3),
\label{eq:SI5_sigma_multimode_rate}
\end{equation}
where
\begin{equation}
\mathcal S_{jk} := -\Tr\!\left[
\mathcal L[R_j]\, \mathcal G[R_k] \right].
\end{equation}

The Fr\'echet derivative $\mathcal G$ is the first derivative of the
matrix logarithm. The bilinear form $(X, Y)\mapsto \operatorname{Tr}\!\left[X\,\mathcal G(Y)\right]$ coincides with the Hessian of the quantum relative entropy evaluated at the full-rank Gibbs state. Consequently,
$-\operatorname{Tr} \left[\mathcal L(\delta\rho)\mathcal G(\delta\rho)\right]$ is the leading quadratic contribution to Spohn's entropy production, which is non-negative for every perturbation $\delta\rho$ of the equilibrium state~\cite{spohn1978}.
Thus, the diagonal contribution $\mathcal S_{11} = -\operatorname{Tr} \left[
\mathcal L[R_1]\, \mathcal G[R_1] \right] \ge0$, with strict positivity for every nontrivial dissipative mode of a primitive Davies generator.

Consequently, we have
\begin{subequations}
\begin{align}
\Sigma(\tau)= & \sum_{j,k\geq1}
c_jc_k^* \frac{e^{(\lambda_j+\overline{\lambda_k})\tau}-1}{\lambda_j+\overline{\lambda_k}}
\mathcal S_{jk} + \mathcal R_{\rm nl}(\tau),
\label{eq:SI5_sigma_integral_general}
\\
\mathcal R_{\rm nl}(\tau)= & \int_0^\tau
O(\|\delta\rho(t)\|^3)\,dt.
\end{align}
\end{subequations}
For a term satisfying $\lambda_j+\overline{\lambda_k}=0$, the corresponding fraction is understood through its continuous limit and equals $\tau$. Equation~\eqref{eq:SI5_sigma_integral_general} is the multimode expression at finite times. In particular, the off-diagonal terms $j\neq k$ account for interference between Liouvillian modes and cannot generally be neglected at short operational times. Separating its diagonal contribution from all other quadratic and nonlinear terms gives
\begin{subequations}
\begin{equation}
\Sigma(\tau) = |c_1|^2 \frac{1-e^{-2\lambda_1\tau}}{2\lambda_1}
\mathcal S_{11} + \mathcal R_{\rm mm}(\tau),
\label{eq:SI5_sigma_with_remainder}
\end{equation}
where
\begin{align}
\mathcal R_{\rm mm}(\tau):= & \sum_{\substack{j,k\geq1\\(j,k)\neq(1,1)}}
c_jc_k^*\frac{e^{(\lambda_j+\overline{\lambda_k})\tau}-1}{\lambda_j+\overline{\lambda_k}}
\mathcal S_{jk} +\mathcal R_{\rm nl}(\tau).
\label{eq:SI5_multimode_remainder}
\end{align}
\end{subequations}

We call the slow-mode approximation controlled over the operational
interval when an explicit estimate
\begin{equation}
|\mathcal R_{\rm mm}(\tau)| \leq \delta_\Sigma(\tau)
\label{eq:SI5_remainder_bound}
\end{equation}
is available and satisfies
\begin{equation}
\delta_\Sigma(\tau)\ll |c_1|^2 \frac{1-e^{-2\lambda_1\tau}}{2\lambda_1} \mathcal S_{11}.
\label{eq:SI5_remainder_small}
\end{equation}
Under these conditions, we have
\begin{equation}
\Sigma(\tau) = |c_1|^2 \frac{1-e^{-2\lambda_1\tau}}{2\lambda_1}
\mathcal S_{11} + O(\delta_\Sigma).
\label{eq:SI5_controlled_single_mode}
\end{equation}

Substituting Eq.~\eqref{eq:SI5_sigma_with_remainder} into the exact heat identity \eqref{eq:SI5_exact_bath_heat} yields
\begin{align}
Q_{\rm bath}(\tau) = k_BT_b \Bigg[& |c_1|^2\frac{1-e^{-2\lambda_1\tau}}{2\lambda_1} \mathcal S_{11} + \mathcal R_{\rm mm}(\tau)- \Delta s(\tau) \Bigg].
\label{eq:SI5_mode_resolved_bath_heat}
\end{align}
This is a mode-resolved approximation to the heat discharged into the bath. It is not, in general, a lower bound, because the sign of the truncated multimode remainder is not fixed without additional assumptions.

\subsection{Comparison of hot and cold preparations}

Consider two preparations
$\rho^{(h)}(0)$ and $\rho^{(c)}(0)$ evolving under the same
$H_f$, bath temperature, and Davies generator. Their respective
operational erasure times are denoted by $\tau_h$ and $\tau_c$.
The exact difference in heat discharged into the bath is
\begin{align}
Q_{\rm bath}^{(h)}(\tau_h)-Q_{\rm bath}^{(c)}(\tau_c)= k_BT_b \Big[
\Sigma^{(h)}(\tau_h)- \Sigma^{(c)}(\tau_c) - \Delta S^{(h)} +\Delta s^{(c)}\Big].
\label{eq:SI5_exact_heat_comparison}
\end{align}
Therefore,
\begin{subequations}
\begin{equation}
Q_{\rm bath}^{(h)}(\tau_h) < Q_{\rm bath}^{(c)}(\tau_c)
\label{eq:SI5_lower_heat_condition}
\end{equation}
holds iff 
\begin{equation}
\Sigma^{(h)}(\tau_h)-\Sigma^{(c)}(\tau_c)
< \Delta S^{(h)}-\Delta S^{(c)}.
\label{eq:SI5_exact_thermodynamic_condition}
\end{equation}
\end{subequations}
Thus, a smaller slow-mode overlap alone does not establish a smaller absolute erasure cost. The entropy changes of the system at the two operational stopping times must also be included.

Within the controlled slow-mode regime, let us define
\begin{subequations}
\begin{equation}
A_\alpha:= |c_1^{(\alpha)}|^2
\frac{1-e^{-2\lambda_1\tau_\alpha}}{2\lambda_1} \mathcal S_{11},
\qquad \text{where} \quad
\alpha\in\{h,c\},
\end{equation}
and 
\begin{equation}
\Sigma^{(\alpha)}(\tau_\alpha)= A_\alpha+\mathcal R_\alpha,
\qquad
|\mathcal R_\alpha|\leq\delta_\alpha.
\end{equation}
\end{subequations}
It follows that
\begin{equation}
\Sigma^{(h)}(\tau_h)-\Sigma^{(c)}(\tau_c)
\leq
A_h-A_c+\delta_h+\delta_c.
\end{equation}
A sufficient condition for a reduced Landauer erasure cost is therefore
\begin{subequations}
\begin{equation}
\Delta S^{(h)}-\Delta S^{(c)} > A_h-A_c+\delta_h+\delta_c.
\label{eq:SI5_sufficient_heat_condition}
\end{equation}
Explicitly,
\begin{align}
\Delta S^{(h)}-\Delta S^{(c)}> \frac{\mathcal S_{11}}{2\lambda_1}
\Big[ |c_1^{(h)}|^2 \bigl(1-e^{-2\lambda_1\tau_h}\bigr)- |c_1^{(c)}|^2 \bigl(1-e^{-2\lambda_1\tau_c}\bigr) \Big]+ \delta_h+\delta_c.
\label{eq:SI5_explicit_sufficient_heat_condition}
\end{align}
\end{subequations}

If both protocols terminate at states with the same von Neumann entropy, or if their entropy changes are independently shown to be equal, condition \eqref{eq:SI5_exact_thermodynamic_condition} reduces to
\begin{equation}
\Sigma^{(h)}(\tau_h) < \Sigma^{(c)}(\tau_c).
\end{equation}
Only under such an additional condition may a reduction of integrated entropy production be directly identified with a reduction of $Q_{\rm bath}$.

\end{document}